# Time-of-flight laser ranging and imaging at 1550 nm using low-jitter superconducting nanowire single-photon detection system

Sijing Chen,[1,2] Dengkuan Liu,[1,2] Wenxing Zhang,[1,2] Lixing You,[1,*] Yuhao He,[1,2] Weijun Zhang,[1] Xiaoyan Yang,[1,2] Guang Wu,[3] Min Ren,[3] Heping Zeng,[3] Zhen Wang,[1] Xiaoming Xie,[1] and Mianheng Jiang[1]

[1]*State Key Laboratory of Functional Materials for Informatics, Shanghai Institute of Microsystem and Information Technology, Chinese Academy of Sciences, 865 Changning Rd., Shanghai 200050, China*

[2]*Graduate University of the Chinese Academy of Sciences, Beijing 100049, PR China*

[3]*State Key Laboratory of Precision Spectroscopy, East China Normal University, Shanghai 200062, China*

*\*Corresponding author: lxyou@mail.sim.ac.cn*

We developed a time-correlated single-photon counting (TCSPC) system based on the low-jitter superconducting nanowire single-photon detection (SNSPD) technology. The causes of jitters in the TCSPC system were analyzed. Owing to the low jitter of the SNSPD technology, a system jitter of 26.8 ps full-width at half-maximum was achieved after optimizing the system. We demonstrated time-of-flight laser ranging at 1550 nm wavelength at a stand-off distance of 115 m, based on this TCSPC system. A depth resolution of 4 mm was achieved directly by locating the centroids of each of the two return signals. Laser imaging was also performed using the TCSPC system. This low-jitter TCSPC system using the SNSPD technology presents great potential in long-range measurements and imaging applications for low-energy-level and eye-safe laser systems.

## 1. Introduction

Time-correlated single-photon counting (TCSPC) is a statistical averaging technique that has single-photon detection sensitivity and picosecond timing accuracy. A time-of-flight system using the TCSPC technique is widely applied in light detection and ranging (LIDAR) instruments [1-4], which are important tools used for measuring the distance between earth and satellites, tracking space debris, etc. Most of the time-of-flight systems operate in the visible wavelength region because of the development of mature laser sources and single-photon detectors (SPDs), such as Si avalanche photodiode (APD) and photomultiplier, which can function at these wavelengths with outstanding performances [5]. However, laser ranging and imaging at 1550 nm are very interesting and have several merits. For example, the 1550 nm wavelength lies in the eye-safe range, is optically invisible, and locates in the atmospheric window. Besides, measurements can be carried out in daylight conditions.

Progress in the fiber communication systems has led to the development of improved laser sources operating at the wavelength of 1550 nm, such as Er-doped lasers. However, imperfections in SPDs operating at 1550 nm limit the time-of-flight applications at this wavelength. This problem can be overcome by using a near-infrared SPD based on InGaAs/InP APD. There are a few reports of laser ranging (and imaging) using time-of-flight systems with InGaAs/InP APDs [6-8]. The depth resolution reached 8 cm due to the timing jitter of 460 ps full-width at half-maximum (FWHM) using InGaAs/InP APD [8]. Owing to the latest InGaAs/InP APD devices with the timing jitter of less than 50 to 70 ps FWHM [9, 10], the depth resolution could be around 1 cm.

In the past decade, a new type of SPD, called superconducting nanowire single-photon detector (SNSPD), has attracted a lot of attention. An SNSPD surpasses an InGaAs/InP APD at near-infrared wavelengths with high quantum efficiency, low dark count rate, high counting rate, low jitter, and wide spectrum sensitivity [11-13]. Warburton et al were the first to report a time-of-flight laser ranging system using SNSPD technology [14]. The system jitter was 70 ps FWHM, which resulted in a surface-to-surface resolution of 1 cm.

In this paper, we present our latest results of time-of-flight laser ranging and imaging at 1550 nm using a low-jitter TCSPC system based on the SNSPD technology. The use of a low-jitter SNSPD and optimization of the TCSPC system effectively reduced the jitter of the TCSPC system to less than 30 ps FWHM. The time-of-flight laser ranging measurement was carried out at a stand-off distance of 115 m, and a surface-to-surface depth resolution of 4 mm was achieved. The 3D imaging on a head sculpture at a stand-off distance of 2.5 m was also demonstrated.

## 2. TCSPC system and its jitter

A typical TCSPC system is mainly composed by a pulsed laser, a TCSPC electronic module, and an SPD and its circuits. The schematic of our TCSPC system using an SNSPD is shown in Fig. 1. A variable optical attenuator is used to attenuate the light power of the pulsed laser source to reach an average photon flux of 1 photon/pulse. The SNSPD device used in the TCSPC system is a 100-nm-wide meander-structured nanowire made of 5 nm-thick NbN film, covering an active area of 15 × 15 μm² with a filling factor of 50%. It is fiber-packaged and mounted on a Gifford–McMahon cryocooler operating at 2.18 K ± 10 mK. The quantum efficiency of the SNSPD at 1550 nm is around 3%, with a dark count rate of 100 Hz when the SNSPD is biased at 96% of its critical current ($I_c$ =34.8 μA). The synchronization trigger pulse of the laser is input to the "Start" port of the classical TCSPC electronic module, and the amplified response pulse from the SNSPD is connected to the "Stop" port. Then the TCSPC electronic module builds up a statistical distribution of the intervals between the "Start" and the "Stop". The distribution typically takes a Gaussian profile, and we can obtain the system jitter from its FWHM.

Each component in the TCSPC system may contribute to the system jitter. For a typical TCSPC system using an SPD with large jitters (for example, >50 ps FWHM), the main contribution of the system jitter originates from the SPD. Influence of other parts (which contribute jitters of <25 ps FWHM) does not play an important role [8]. However, in a TCSPC system using a low-jitter SPD, such as a SNSPD, the contribution of jitter from other parts may be non-negligible. Therefore, it is necessary to analyze the jitter of each component systematically and optimize the system by choosing low-jitter components. Summarizing the possible causes of the system jitter, the total jitter of the TCSPC system $j_{total}$ may be expressed as given in Eq. (1)

$$j_{total} = \sqrt{j_{SNSPD}^2 + j_{TCSPC}^2 + j_{laser}^2 + j_{sync}^2} \quad (1)$$

where $j_{SNSPD}$ is the jitter of the SNSPD output pulse signal; it is composed of the jitters originating from the device and its circuit, which are somehow correlated and difficult to separate. The value of the jitter is typically a few tens of picoseconds. It cannot be determined until we know $j_{total}$ and jitters of all the other parts. $j_{TCSPC}$ is the jitter of the TCSPC module, which is determined by the commercial TCSPC electronic module itself; $j_{laser}$ and $j_{sync}$ are the jitters from the laser source and its synchronization signal, respectively, which are also known from the specification of the laser. The jitter induced by the dispersion in the several-meters-long single-mode fiber in the system is typically no more than one picosecond, which can be neglected [15].

Based on the analysis above, each component of the TCSPC system was carefully chosen to lower the system jitter. We employed a mode-locked femtosecond-pulsed fiber laser (FPL-01CAF from Calmar Inc.) centered at 1550 nm, with a spectrum width of 5 nm and a repetition rate of 20 MHz. The jitters of the pulsed laser and its synchronization trigger were 60 fs and 4 ps, respectively. Further, a low-jitter TCSPC module (SPC-130 board card from Becker & Hickl GmbH) was used to replace the popular TCSPC module PicoHarp 300 (from PicoQuant GmbH), which is widely used in other TCSPC systems [8, 16, 17]. The former has a jitter of 7.6 ps FWHM whereas the latter has a jitter of 26.1 ps FWHM. The width of the time bin per channel for SPC-130 and PicoHarp 300 is 813 fs and 4 ps respectively. The choice of the amplifier is also very sensitive. We compared two amplifiers. One was the low-noise amplifier LNA-650 from RF Bay Inc., with a gain of 50 dB and a bandwidth of 30 KHz-600 MHz; the other was the Phillips 6954 with a gain of 40 dB and a bandwidth from 100 kHz to 1.8 GHz. The system jitter using the former amplifier is obviously smaller than the value using the latter. The possible reason is that a poor impendence match was observed from the amplified pulse signal using Phillips 6954. On the other hand, an increased signal-noise-ratio may also help to reduce the jitter due to a larger gain of LNA-650.

By choosing the suitable components, the jitter of our TCSPC system is effectively reduced. Fig. 2(a) shows the system jitter $j_{total}$ when SNSPD is biased at $0.96I_c$. By Gaussian fitting, we have $j_{total}$=26.8 ps FWHM. This is one of lowest jitters reported for a TCSPC system operated at 1550 nm wavelength, which is also lower than the jitter of the system (37 ps FWHM) using an SNSPD with a superconducting single-flux quantum readout circuit [18]. Such a low-jitter TCSPC system will improve the depth resolution of the time-of-flight laser ranging and imaging application effectively. From formula (1) we can derive that the value of the jitter $j_{SNSPD}$ is 25.4 ps FWHM, which is among the lowest reported values for an SNSPD [17,19-22]. A latest article reported a novel SNSPD atop nanophotonic waveguides with a lower $j_{SNSPD}$ of 18 ps FWHM, which is credited to a better amplifier with the wider bandwidth of 15 GHz [17].

Although the origin of the intrinsic jitter of the SNSPD is still not well understood, there are some encouraging measurement results that can help understand the mechanism further. The blue dots and red squares in Fig. 2(b) show the system jitter $j_{total}$ and the calculated $j_{SNSPD}$ at different $I_b$. We noticed that the jitter decreases as the $I_b$ increases. Similar results were also reported in reference [18]. In order to judge whether the jitter depends on $I_b$ or $I_b/I_c$, we changed $I_c$ of SNSPD by altering its operating temperature. The result indicates that the jitter keeps unchanged with the same $I_b$, so $j_{SNSPD}$ is determined by $I_b$ rather than $I_b/I_c$. Therefore, the devices with a higher critical current can be biased at a higher $I_b$ then have a lower jitter. This can be realized using either thicker thin film [23] or a wider nanowire. The disadvantage is a lower quantum efficiency of SNSPD. However, with a higher laser power it is possible to work with a lower QE SNSPD in the TCSPC applications.

## 3. Time-of-flight laser ranging and imaging at 1550 nm

A laser ranging experiment was performed using the time-of-flight TCSPC technique. The experimental platform (shown in Fig. 3) included the TCSPC system that had a jitter of 26.8 ps FWHM. The average output power of the laser was 2 mW and the repetition rate was 20 MHz, corresponding to the single-pulse energy of 0.1 nJ. The optical transceiver system was operated in a coaxial mode using two reflective mirrors. The laser, after being passed through a single-mode fiber, was expanded and collimated by a reflective collimator. The diameter of the output beam was 6 mm, which increased to about 36 mm at the target located 115 m away. A reflection Newton telescope (aperture diameter 140 mm) was used to receive the retro-reflected photons. An optical band pass interference filter (OBPF) centered at 1550 nm was used to block the stray daylight before the retro-reflected photons were further converged by a lens. Then the photons were coupled to a multimode fiber having a core size of 50 μm. Since the SNSPD was optically coupled through a single-mode fiber, the multimode fiber was connected directly to the single-mode fiber by an FC–PC connector. Optical loss from the telescope to the SNSPD was about 6 dB, which can be reduced further in the future. Note that an SPC-130 system is designed to work in the reversed start–stop mode which is more suitable for high repetition rate operation. The amplified response pulse from the SNSPD was connected to the "CFD" port as a "start", while the synchronization trigger pulse of the laser was input to the "SYNC" port as a "stop" [24].

In order to identify the surface-to-surface separation, two identical mirrors were attached to a translation stage and mounted on an opto-mechanical rail. The rail was placed at a distance of approximately 115 m from the telescope. These mirrors could act as two closely spaced reflecting surfaces with variable separations of up to 50 mm.

Four typical photon-return histograms were recorded by SPC-130, which are shown in Fig. 4. The acquisition time was adjusted to assure that the maximum counts of each case reached nearly $10^5$. Figs. 4(a-d) correspond to the separations of the retro-reflected mirrors of 10, 4, 3.5, and 3 mm, respectively. The blue solid lines and red dashed lines correspond to the two-peak Gaussian fittings of the time correlation for M3 and M4, respectively. By locating the centroids of each of the two return signals, we could obtain the separations between them. The distance was calculated directly using the following formula: $L = c \times \Delta t/2$, where $c$ is the speed of light in air and $\Delta t$ is the time interval of the two centroids of the return signals. When M3 and M4 were separated by 10 mm, the two peaks of the histograms were clearly separated by a time interval of 66.7 ps, as shown in Fig. 4(a). For separations of 4 and 3.5 mm between the mirrors M3 and M4, although the two peaks overlapped, we could still differentiate them easily (Fig. 4(b) and (c)) using the two-peak Gaussian fittings. However, when the separation was further reduced to 3 mm, two peaks could not be distinguished. Thus, the depth resolution obtained was better than 4 mm, which corresponded to a 26.7 ps time-of-flight of photons. This value is consistent with the 26.8 ps FWHM jitter of the laser ranging system. For a separation of less than 3.5 mm, an optimized signal-processing algorithm based on a reversible jump Markov chain Monte-Carlo method may provide a better depth resolution [14].

Preliminary laser imaging at a stand-off distance of 2.5 m was also demonstrated. Another SNSPD detector was used in the system which produced the system jitter of 31 ps FWHM. The two mirrors in Fig. 3 were replaced with a 35 cm × 40 cm head sculpture (shown in Fig. 5(a)). The head sculpture was scanned in the Y–Z plane with a step width of 1 cm. We scanned 29 × 35 pixels on the head sculpture, and the depth information of each pixel was recorded by the TCSPC system. The acquisition time for each pixel was 10 s. A 3D shaded surface image was produced from the data using "Matlab", which is shown in Fig. 5(b). The profile of the head sculpture can be identified easily from the laser image. A depth difference of 5 mm between the lower lip and the chin can be perceived in the laser image. Space resolution can be improved further by decreasing the laser beam size and the scanning step, while the depth resolution can be improved by decreasing the jitter of the TCSPC system.

## 4. Conclusion

We developed a TCSPC system with a jitter of 26.8 ps FWHM using a low-jitter SNSPD. Based on the TCSPC system, time-of-flight laser ranging at 1550 nm wavelength at a stand-off distance of 115 m was performed. A direct surface-to-surface resolution of less than 4 mm was achieved, which is, to our knowledge, the best result reported to date at the wavelength of 1550 nm. Laser imaging at a stand-off distance of 2.5 m was also demonstrated. In the future, by improving the SNSPD devices as well as further optimizing the TCSPC system, we believe that the jitter down to less than 15 ps FWHM can be realized. The low-jitter TCSPC system using an SNSPD will be of great potential in LIDAR applications at 1550 nm wavelength.

## 5. Acknowledgments

Sijing Chen and Dengkuan Liu contributed equally to this work. We acknowledge financial support from the National Natural Science Foundation of China (Grant No. 91121022), 973 Program (Grant No. 2011CBA00202), 863 Program (Grant No. 2011AA010802), and the "Strategic Priority Research Program (B)" of the Chinese Academy of Sciences (Grant Nos. XDB04010200 and XDB04020100).

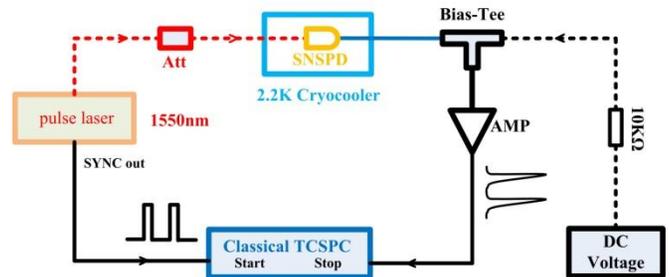

Fig. 1. Schematic of a typical TCSPC system based on SNSPD technology. The red dashed lines represent the route of light; the black dashed lines represent the route of the dc bias. The blue straight line represents the route of both input dc bias and output response pulse from the SNSPD. ATT: optic attenuator; SYNC:

synchronizing signal.

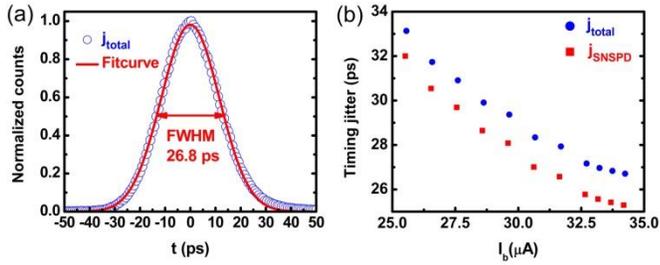

Fig. 2. (a) Measured jitter of our TCSPC system (blue circles) and Gaussian fit curve for it (red solid line). The bias current is set at 0.96 $I_c$. (b) Dependence of $j_{total}$ (blue dots) and $j_{SNSPD}$ (red squares) on $I_b$. All the jitters are FWHM values.

Fig. 3. Schematic of the laser ranging system based on our optimized TCSPC system that uses the low-jitter SNSPD. M1, M2: 1-inch round protected gold mirrors; M3, M4: gold-plated mirrors; OBPF: optical band-pass filter (center wavelength: 1550 nm, FWHM: 6.47 nm); f: convex lens with 3 cm focal length; MMF: multimode fiber; SMF: single-mode fiber.

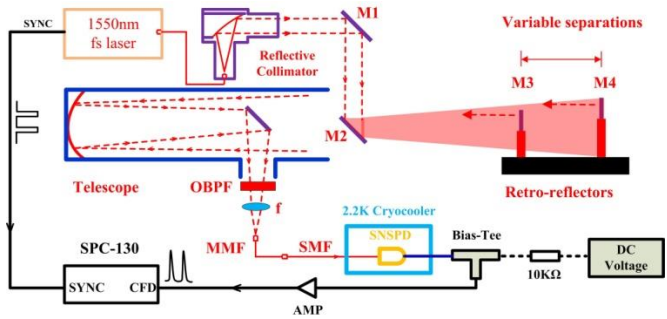

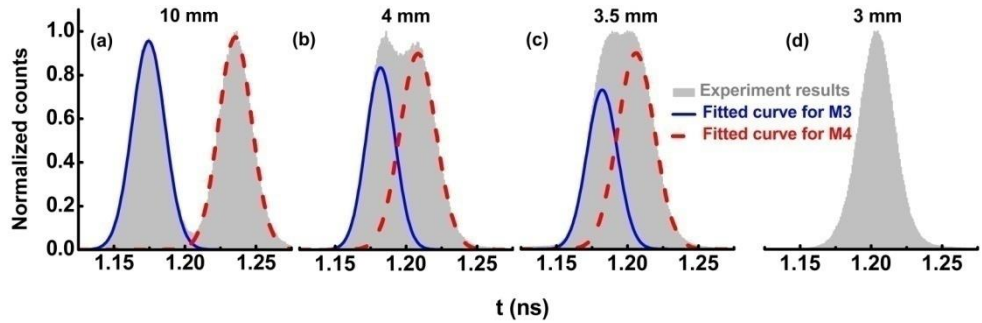

Fig. 4. TCSPC results when the retro-reflected mirrors were separated by 10, 4, 3.5, and 3 mm, respectively.

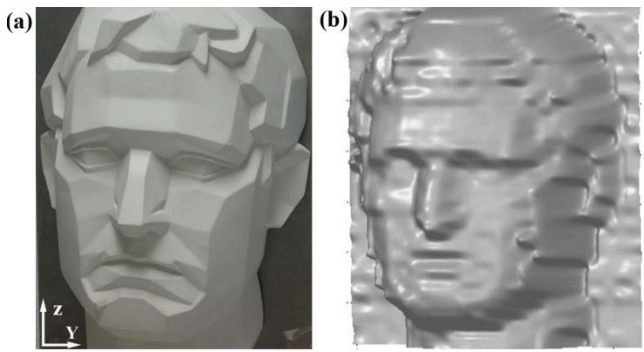

Fig. 5. (a) Photograph of the plaster cast. (b) 3D shaded surface image created by "Matlab" using data (29 × 35 pixels) from the TCSPC measurements.